\documentclass[prl,nofootinbib,preprint,superscriptaddress]{revtex4}

\usepackage{graphicx}
\usepackage{amsmath}
\usepackage{amsfonts}
\usepackage{amssymb}
\usepackage{dsfont}
\usepackage{dcolumn}
\usepackage{bm}
\usepackage{amsmath,amssymb}
\usepackage{graphicx}
\usepackage[pdftex]{hyperref}
\usepackage{pstricks}
\usepackage{color}

\newcommand{\beqn}{\begin{eqnarray}}
\newcommand{\eeqn}{\end{eqnarray}}
\newcommand{\be}{\begin{equation}}
\newcommand{\ee}{\end{equation}}

\newcommand{\mathsym}[1]{{}}

\begin{document}

\title{$f(T)$ gravity from holographic Ricci dark energy model with new boundary conditions}

\author{Peng Huang}

\affiliation{Institute of Theoretical Physics, Beijing University of Technology\\ Beijing 100124, China}
\author{Yong-Chang Huang}

\affiliation{Institute of Theoretical Physics, Beijing University of Technology\\ Beijing 100124, China}
\affiliation{Kavli Institute for Theoretical Physics, Chinese Academy of Sciences\\ Beijing 100080, China}
\affiliation{CCAST (World Lab.), P.O. Box 8730\\ Beijing 100080, China}
\author{Fang-Fang Yuan}

\affiliation{Institute of Theoretical Physics, Beijing University of Technology\\ Beijing 100124, China}




\begin{abstract}

Commonly used boundary conditions in reconstructing $f(T)$ gravity from holographic Ricci dark energy model (RDE) are found to cause some problem, we therefore propose new boundary conditions in this paper. By reconstructing $f(T)$ gravity from the RDE with these new boundary conditions, we show that the new ones are better than the present commonly used ones since they can give the physically expected information, which is lost when the commonly used ones are taken in the reconstruction, of the resulting $f(T)$ theory. Thus, the new boundary conditions proposed here are more suitable for the reconstruction of $f(T)$ gravity.


\end{abstract}

\maketitle

\section{1. Introduction}
\label{sec:intro}

Cosmological observations \cite{SN,riess,Spergel,Komatsu,T,j,D,B} denote that the universe is accelerating, this accelerative phenomenon implies two possibilities in theory of gravity and cosmology: the theory of gravity should be modified in cosmological scale, or there is an exotic energy component with repulsive gravity. Correspondingly, two different approaches are developed to explain the acceleration. The first is to modify the theory of gravity, and the second is to introduce a new energy form, named dark energy, with an appropriate index of equation of state (EOS) $\omega$ to trigger the acceleration.

These two approaches are very different, they are developed with completely different motivations. However, when dark energy is geometrically defined, the common geometrical character of these two approaches can relate them to each other tightly. This is what happens when dark energy is described by holographic dark energy models, in which the dark energy is defined by geometrical objects of spacetime, such as event horizon $R_h$ (holographic dark energy model \cite{li}\cite{li2}), cosmological conformal time $\eta$ (agegraphic dark energy model \cite{cai1}\cite{cai2}), and the Ricci scalar $R$ (holographic Ricci dark energy model \cite{gao}).

Recently, lots of efforts have been done in reconstructing modified theory of gravity, such as $f(R)$ and $f(T)$, from holographic dark energy models \cite{Bamba:2012vg,Feng:2008hk,Feng:2008kz,Karami:2010bu,Karami:2010xy,18,19,20,Huang:2013una}. The advantage of doing reconstruction like this is the resulting theory will accommodate a accelerated universe automatically. The methodology in these reconstructions can be divided in to three steps: (1) get the modified Friedmann equation form the modified theory of gravity. (2) express the density and pressure of dark energy in modified Friedmann equation as function of the Lagrangian density of the original theory of gravity, such as Ricci scaler $R$ for general relativity and torsion scaler $T$ for Teleparallel Equivalent of General Relativity. This step will lead to a differential equation. (3) solve the differential equation with appropriate boundary conditions and get the final form of the modified theory of gravity.

Apparently, the choice of the boundary condition is crucial for the determination of the final form of the modified theory of gravity. The present commonly used boundary conditions are always set the boundary on the birth time of the universe. In this paper, we will point out that these boundary conditions will cause some problem. We will also propose new boundary conditions and proof that, by exhibiting the hole process of reconstructing $f(T)$ gravity from RDE model, the new boundary conditions are indeed more suitable for reconstruction of modified theory.

The paper is organized as follows: in Sec. 2, we briefly review the problem caused by present commonly used boundary conditions; in Sec. 3, we reconstruct $f(T)$ gravity according to RDE model with new boundary conditions and show the advantage of these boundary conditions. Finally, summary and discussion are given in Sec.4.

\section{2. The problem of the commonly used boundary conditions}

Einstein's general theory of relativity (GR) can be rewritten in teleparallel language. The resulting theory is known as the Teleparallel Equivalent of General Relativity (TEGR) \cite{tegr,Hayashi,Pereira,Kleinert,Sonester}. In order to explain both the early inflation and the late time acceleration of the universe, the teleparallel Lagrangian density also has been extended to a function of $T$ as $f(T)$ \cite{10}\cite{11}, similar to the spirit of extending from $R$ to $f(R)$.

The corresponding Friedmann equations in the flat spatial Friedmann-Robertson-Walker (FRW) universe of $f(T)$ gravity are given by
\begin{equation}
12H^2f_T(T)+f(T)=2\rho,\label{fT1}
\end{equation}
\begin{equation}
48H^2\dot{H}f_{TT}(T)-(12H^2+4\dot{H})f_T(T)-f(T)=2p.\label{fT2}
\end{equation}
One can decompose $f(T)$ into $f(T)=g(T)+T$ as that in \cite{18}, where $T$ corresponds to the original TEGR and $g(T)$ denotes the departure of $f(T)$ gravity from TEGR. Then, Eq.(\ref{fT1}) and Eq.(\ref{fT2}) turn into
\begin{equation}
3H^2=\rho-\frac{1}{2}g-6H^2g_T,\label{ft3}
\end{equation}
\begin{equation}
-3H^2-2\dot{H}=p+\frac{1}{2}g+2(3H^2+\dot{H})g_T-24\dot{H}H^2g_{TT},\label{ft4}
\end{equation}
Comparing Eq.(\ref{ft3}) and (\ref{ft4}) with ordinary Friedmann equations, one gets relation between $\rho_D$, $p_D$ and $g(T)$ that \cite{18}
\begin{equation}
\rho_D=-\frac{1}{2}g-6H^2g_T,\label{gt1}
\end{equation}
\begin{equation}
p_D=\frac{1}{2}g+2(3H^2+\dot H)g_T-24\dot H H^2g_{TT},\label{gt2}
\end{equation}
which can be combined into
\begin{equation}
\rho_D+p_D=\rho_D(1+\omega_D)=2\dot H g_T-24\dot H H^2g_{TT}.\label{gt3}
\end{equation}
We can see from Eq.(\ref{gt3}) that one can solve this differential equation and thus get the concrete form of $f(T)$ gravity when the density and pressure of dark energy is define by function of the scalar torsion, $T$.

The dark energy density in RDE model is defined by
\begin{equation}
\rho_D=3\alpha(2H^2+\dot H).\label{rdeenergydensity}
\end{equation}

The parameter $\alpha$ is crucial in determining the evolutionary
behavior of RDE. According to the observational constraints from the joint
analysis of data of SN, BAO, and WMAP5 in \cite{rip2}, the best-fit result for $\alpha$ is $\alpha=0.359^{+0.024}_{-0.025}$, which shows that RDE will more likely behave as a phantom energy, it is expected that the big rip \cite{rip} will occur in a finite time.

In matter dominated era, since $H=\frac{2}{3t}$ and using the well know result in TEGR that $T=-6H^2$, $\dot H$ can be rewrited as $\dot H=\frac{T}{4}$, thus the Ricci dark energy density can be described by function of torsion scaler as
\begin{equation}
\rho_D=-\frac{\alpha T}{4}.
\end{equation}
The value of $\omega_D$ can be get by inserting expression of $\rho_D$ into continuity equation, which leads to
\begin{equation}
(4H\dot H + \ddot H)+3H(1+\omega_D)(2H^2+\dot H)=0.\label{conteq2}
\end{equation}
Then $\omega_D$ can be get by inserting $H=\frac{2}{3t}$ into Eq.(\ref{conteq2}), the result is
\begin{equation}
\omega_D=0,
\end{equation}
which tells that the dark energy behaves the same as that of matter during the matter dominated era. In fact, since $\rho_D=3\alpha(2H^2+\dot H)\propto H^2$ in this special time duration, the same result can be get through similar argument as that in \cite{shu}. Now, one know that Eq.(\ref{gt3}) turns into
\begin{equation}
T^2g_{TT}+\frac{T}{2}g_T+\frac{\alpha T}{4}=0,
\end{equation}
whose solution is
\begin{equation}
g(T)=2C_1\sqrt {T}-\frac{1}{2}\alpha T+C_2,
\end{equation}
then, the form of $f(T)$ is
\begin{equation}
f(T)=(1-\frac{1}{2}\alpha)T+2C_1\sqrt {T}+C_2.\label{rdematter}
\end{equation}
By using the common used boundary conditions that \cite{Capozziello:2005ku,18,Feng:2008kz,Huang:2013una}
\begin{equation}
f(T)_{t=0}=T_0,\label{old1}
\end{equation}
\begin{equation}
(\frac{df(T)}{dt})_{t=0}=(\frac{dT}{dt})_{t=0},\label{old2}
\end{equation}
one then gets
\begin{equation}
f(T)=(1-\frac{\alpha}{2})T+\alpha\sqrt {T_0T}-\frac{\alpha}{2}T_0.\label{rdematterfinal}
\end{equation}

Now we can see what the problem is. Since $\omega_D=0$ for the dark energy in matter dominated cosmological era, the dark energy behaves the same as that of matter, thus, this obviously leads to a rescaling of the right hand of the Friedmann equation, and correspondingly, $f(T)$ gravity should also be a rescaling of the original TEGR, that is $h(\alpha)T$, $h(\alpha)$ here is a parameter which is a function of $\alpha$ in RDE. However, such important expected information is lost in Eq.(\ref{rdematterfinal}), which denotes that the present commonly used boundary conditions are not adequate enough to give the physically expected, thus, new boundary conditions are needed.

\section{3. $f(T)$ gravity from RDE model with new boundary conditions}

In the radiation dominated cosmological era, $a(t)=bt^{\frac{1}{2}}$ ($b$ is a positive constant) and $H=\frac{1}{2t}$, it's easy to find that $\rho_D=3\alpha(2H^2+\dot H)=0$. Then, Eq.(\ref{gt3}) turns into
\begin{equation}
Tg_{TT}+\frac{1}{2}g_T=0,
\end{equation}
Again, by using the commonly used boundary conditions that
\begin{equation}
f(T)_{t=0}=T_0,
\end{equation}
\begin{equation}
(\frac{df(T)}{dt})_{t=0}=(\frac{dT}{dt})_{t=0},
\end{equation}
we can get the final form of $f(T)$ gravity in radiation dominated epoch that
\begin{equation}
f(T)=T.\label{tradiation}
\end{equation}

This result says that $f(T)$ gravity reconstructed from RDE model in radiation dominated era is just the TEGR itself, which is equivalent to general relativity, no modification happens. This is consistent with the fact that the RDE dark energy density is zero in this cosmological era, thus, it has no impact on the corresponding gravity theory. This result also denotes that the boundary conditions used in this process give the expected resulting $f(T)$ gravity precisely, thus, these boundary conditions are adequate enough for the reconstruction in radiation dominated era.

In the matter dominated cosmological era, it has been expatiated in Sec.2 that the reconstructed $f(T)$ gravity has the form
\begin{equation}
f(T)=(1-\frac{1}{2}\alpha)T+2C_1\sqrt {T}+C_2.\label{rdematter}
\end{equation}
In order to decide what the exact $f(T)$ gravity is, the two integral constant, $C_1$ and $C_2$, must be fixed by appropriate boundary conditions. The present commonly used boundary have been proofed to cause some problem and new boundary conditions are needed.

However, one should notice the fact that the present boundary conditions do well in the radiation dominated cosmological era, which denotes that they are indeed the correct ones for the reconstruction in this special cosmological era. Thus, new boundary conditions must be based on these old boundary conditions, and at the same time, have some improvement on the details of the old ones, otherwise, the new boundary conditions cannot be able to trace back to the present boundary conditions which work well in radiation dominated cosmological era.

Under this consideration, it is natural to require that new boundary conditions which distinguish themselves from the old ones are the new time point on which they are defined. For matter dominated cosmological era, its boundary conditions should defined on the critical time which is the point of time that the radiation dominated era turns into matter dominated era, i.e.,
\begin{equation}
f(T_M)_{t=t_{R\rightarrow M}}=f(T_R),\label{new1}
\end{equation}
\begin{equation}
(\frac{df(T_M)}{dt})_{t=t_{R\rightarrow M}}=(\frac{df(T_R)}{dt})_{t=t_{R\rightarrow M}},\label{new2}
\end{equation}
where $T_M$ and $T_R$ denote the torsion scaler of the $f(T)$ gravity in radiation and matter dominated cosmological era, respectively; $t_{R\rightarrow M}$ denotes the critical time when radiation dominated era changes into matter dominated era.

It has been shown in Sec.2 that the old boundary conditions, defined as Eq.(\ref{old1}) and (\ref{old2}), lose the key expected character of the resulting theory: the rescaling of the original TEGR. Thus, only when the resulting $f(T)$ gravity reconstructed with these new boundary conditions, defined as Eq.(\ref{new1}) and (\ref{new2}), shows the expected rescaling character, then can we claim that the new boundary conditions are better than the old ones.

To see what result these new boundary conditions will lead to, we use Eq.(\ref{tradiation}) to rewrite Eq.(\ref{new1}) and Eq.(\ref{new2}) into
\begin{equation}
f(T_M)_{t=t_{R\rightarrow M}}=T_R,\label{new11}
\end{equation}
\begin{equation}
(\frac{df(T_M)}{dt})_{t=t_{R\rightarrow M}}=(\frac{dT_R}{dt})_{t=t_{R\rightarrow M}}.\label{new22}
\end{equation}
Eq.({\ref{new22}}) can further be rewritten as
\begin{equation}
(\frac{df(T_M)}{dT_M}\frac{dT_M}{dt})_{t=t_{R\rightarrow M}}=(\frac{dT_R}{dt})_{t=t_{R\rightarrow M}}.\label{new1234}
\end{equation}
In radiation dominated era, $H=\frac{1}{2t}$; in matter dominated era, $H=\frac{2}{3t}$. Using $T=-6H^2$, one can get
\begin{equation}
T_M=\frac{16}{9}T_R.\label{tmtr}
\end{equation}
With Eq.(\ref{tmtr}), Eq.(\ref{new1234}) turns into
\begin{equation}
(\frac{df(T_M)}{dT_M})_{t=t_{R\rightarrow M}}=\frac{9}{16}.\label{new222}
\end{equation}
Combing Eq.(\ref{new11}), Eq.(\ref{new222}) and Eq.(\ref{rdematterfinal}), we can fix the integral constant $C_1$ and $C_2$ as
\begin{equation}
C_1=\frac{8\alpha-7}{12}T_R^\frac{1}{2},
\end{equation}
\begin{equation}
C_2=-\frac{8\alpha-7}{9}T_R.
\end{equation}
Inserting these integral constant back into Eq.(\ref{rdematterfinal}), one can find that the result is $f(T_M)=\frac{9}{16}T_M$, which says that, in the matter dominated cosmological era, the $f(T)$ gravity reconstructed from RDE model with new boundary condition is
\begin{equation}
f(T)=\frac{9}{16}T.\label{newresult}
\end{equation}
$f(T)$ gravity in this form indeed shows the rescaling of the original theory, thus, the new boundary conditions proposed in this paper are indeed better than old ones, since they give the expected information which is lost when the old boundary conditions are used.

In the dark energy dominated cosmological era, it has been showed in \cite{Huang:2013una} that the reconstructed $f(T)$ gravity takes the form as
\begin{equation}
f(T)=(1-2\alpha-\frac{\alpha}{h})T+2C_3\sqrt T+C_4.\label{rdede}
\end{equation}
The boundary conditions used in \cite{Huang:2013una} to fix the integral constants are Eq.(\ref{old1}) and Eq.(\ref{old2}), and the final form of the $f(T)$ gravity is
\begin{equation}
f(T)=(1-2\alpha-\frac{\alpha}{h})T+(4\alpha+\frac{2\alpha}{h})\sqrt {T_0T}+(2\alpha+\frac{\alpha}{h})T_0.
\end{equation}

However, we have shown that these commonly used boundary conditions cause problem in the reconstruction in matter dominated era. On contrary, new boundary conditions that set on the critical time are better than the old ones. It's indeed more reasonable to set the boundary conditions on critical time, on which the radiation dominated era turns into matter dominated era, for the boundary time of a matter dominated cosmological era apparently is the critical time ($t=t_{R\rightarrow M}$) but not the birth time of the universe ($t=0$).

For the same reason, in dark energy dominated era, we also need to set the boundary on the critical time when matter dominated era turns into dark energy dominated era. Denoting this critical time as $t=t_{M\rightarrow D}$, and $T_D$ the torsion scaler in dark energy dominated cosmological era, we have two new boundary conditions for the reconstruction of $f(T)$ gravity in dark energy dominated era that
\begin{equation}
f(T_D)_{t=t_{M\rightarrow D}}=f(T_M),\label{new3}
\end{equation}
\begin{equation}
(\frac{df(T_D)}{dt})_{t=t_{M\rightarrow D}}=(\frac{df(T_M)}{dt})_{t=t_{M\rightarrow D}},\label{new4}
\end{equation}

Using Eq.(\ref{newresult}), we can rewrite new boundary conditions defined by Eq.(\ref{new3}) and Eq.(\ref{new4}) into
\begin{equation}
f(T_D)_{t=t_{M\rightarrow D}}=\frac{9}{16}T_M,\label{new33}
\end{equation}
\begin{equation}
(\frac{df(T_D)}{dt})_{t=t_{M\rightarrow D}}=\frac{9}{16}(\frac{dT_M}{dt})_{t=t_{M\rightarrow D}},\label{new44}
\end{equation}
For convenience of calculation, we can rewrite Eq.({\ref{new44}}) as
\begin{equation}
(\frac{df(T_D)}{dT_D}\frac{dT_D}{dt})_{t=t_{R\rightarrow M}}=\frac{9}{16}(\frac{dT_M}{dt})_{t=t_{M\rightarrow D}}.\label{new444}
\end{equation}
In matter dominated era, $H=\frac{2}{3t}$; in dark energy dominated era for RDE model, because of its Phantom character \cite{rip2}, a Hubble parameter with future singularity should be assumed. In this paper, we take the same assumption as that in \cite{bamba} with
\begin{equation}
H=h(t_s-t)^{-1},\label{rdeh}
\end{equation}
then, by using of $T=-6H^2$, one can get
\begin{equation}
T_D=\frac{-6h^2}{t_S^2-\frac{8}{3T_M}-8t_S(-6T_M)^{-\frac{1}{2}}}.
\end{equation}
After a similar series of calculation, we can get the final form of the $f(T)$ gravity reconstructed from RDE model in dark energy dominated era with new boundary conditions as
\begin{equation}
f(T)=\frac{3}{2}(t_S-\frac{6h}{\sqrt {-6T}})^{-2}.\label{final}
\end{equation}

\section{4. Summary and discussion}

In this paper, we reconstruct $f(T)$ gravity according to holographic Ricci dark energy model with new boundary conditions. We review the reconstruction done in the matter dominated era first, by using the commonly used boundary conditions, and show that this have caused problem since they cannot give the expected rescaling character of the original theory, in other words, the commonly used boundary conditions cannot give the physically expected result, thus, new boundary conditions are needed.

Notice the fact that the old boundary conditions do well in radiation dominated era while lose its effectiveness in matter dominated era, we then realize that we must be more cautious about on what time the boundary conditions are defined. For radiation dominated era, it is indeed natural to define its boundary time on the birth time of the universe ($t=0$); however, for matter dominated era, the boundary time should be chosen on the critical time when radiation dominated era switches to matter dominated era ($t=t_{R\rightarrow M}$), similarly, for dark energy dominated era, the boundary time should be set on the critical time when matter dominated era switches to dark energy dominated era ($t=t_{M\rightarrow D}$), otherwise, the boundary condition will too vague to give the expected result.

Under this these consideration, we propose new boundary conditions that
\begin{equation}
f(T_J)_{t=t_{I\rightarrow J}}=f(T_I),
\end{equation}
\begin{equation}
(\frac{df(T_J)}{dt})_{t=t_{I\rightarrow J}}=(\frac{df(T_I)}{dt})_{t=t_{I\rightarrow J}},
\end{equation}
the Latin subindexes $I$ and $J$ denote two adjacent cosmological eras which have different energy component in dominated position, and on the critical time, cosmological era indexed by $I$ switches into cosmological era indexed by $J$. Concretely speaking, for matter dominated era, the new boundary conditions for $f(T)$ gravity reconstruction is
\begin{equation}
f(T_M)_{t=t_{R\rightarrow M}}=f(T_R),
\end{equation}
\begin{equation}
(\frac{df(T_M)}{dt})_{t=t_{R\rightarrow M}}=(\frac{df(T_R)}{dt})_{t=t_{IR\rightarrow M}},
\end{equation}
and for dark energy dominated era, the new boundary conditions are
\begin{equation}
f(T_D)_{t=t_{M\rightarrow D}}=f(T_M),
\end{equation}
\begin{equation}
(\frac{df(T_D)}{dt})_{t=t_{M\rightarrow D}}=(\frac{df(T_M)}{dt})_{t=t_{M\rightarrow D}}.
\end{equation}
These new boundary conditions have advantage over old ones in the sense that they give the resulting theory with more information which is physically expected, it's in this sense that these new boundary conditions are indeed more suitable for corresponding reconstruction of $f(T)$ gravity. It should be noted that the form of the solution got by original boundary conditions, Eq.(\ref{rdede}), is consistent with solutions got in \cite{18}\cite{Jamil:2012nma}, and Eq.(\ref{final}) can been regarded as an improvement of this solution by utilizing new boundary conditions. Thus, it's valuable to investigate how much improvement do these new boundary conditions bring about comparing with the old one. For the same reason, it's also interesting to study the difference between the solution got in this paper by analytic calculation and solution got by cosmographic reconstruction \cite{Aviles:2013nga}. Furthermore, it's well known that there are four different types of future singularities of the universe (Big Rip or Cosmic Doomsday, see \cite{Caldwell:2003vq} \cite{Nojiri:2003vn} \cite{Nojiri:2005sx} for detail), the singularity investigated (Eq.(\ref{rdede})) in this work belongs to Type I. Systematic study of future singularities in $f(T)$ theories can been found in \cite{Bamba:2012vg}\cite{Bamba:2012cp} and references there in. Correspondingly, it's also interesting to see what happens when the form of Eq.(\ref{rdede}) is replaced with other types of future singularities.

\acknowledgments

The work is partly supported by National Natural Science Foundation of China (No. 11275017 and No. 11173028).


\end{document}